\documentclass[aps,floats,prb,psfig,epsf,showpacs,twocolumn,superscriptaddress]{revtex4}

\usepackage{amsfonts}
\usepackage{amsmath}
\usepackage{amssymb}
\usepackage{graphicx}
\usepackage[dvipsnames]{xcolor}
\def\a{$\alpha$~}
\def\g{$\gamma$~}
\def\ag{\mbox{$\alpha$-$\gamma$} }

\begin{document}

% \preprint{HEP/123-qed}
\title{Nonlocal correlations in the vicinity of the $\alpha$-$\gamma$ phase transition in iron\\ within a DMFT plus spin-fermion model approach}

\author{A. A. Katanin}
\affiliation{Miheev Institute of Metal Physics, Russian Academy of Sciences, 620137 Yekaterinburg, Russia}
\affiliation{Ural Federal University, 620002 Yekaterinburg, Russia}

\author{A. S. Belozerov}
\affiliation{Miheev Institute of Metal Physics, Russian Academy of Sciences, 620137 Yekaterinburg, Russia}
\affiliation{Ural Federal University, 620002 Yekaterinburg, Russia}

\author{V. I. Anisimov}
\affiliation{Miheev Institute of Metal Physics, Russian Academy of Sciences, 620137 Yekaterinburg, Russia}
\affiliation{Ural Federal University, 620002 Yekaterinburg, Russia}

\date{\today}

\begin{abstract}
We consider nonlocal correlations in iron in the vicinity of the $\alpha$-$\gamma$ phase transition within the spin-rotationally-invariant dynamical mean-field theory (DMFT) approach,
%[Phys. Rev. B \textbf{87}, 125138 (2013)]
combined with the  recently proposed spin-fermion model of iron. %[Phys. Rev. B \textbf{91}, 195123 (2015)]. 
The obtained nonlocal corrections to DMFT 
%to the energy of $\alpha$ phase 
yield a decrease of the Curie temperature of the $\alpha$ phase, leading to an agreement with its experimental value. 
%very close to its experimental value, and the
We show that the corresponding nonlocal corrections to the energy of the $\alpha$ phase are crucially important to obtain the proximity of energies of $\alpha$ and $\gamma$ phases in the vicinity of the iron $\alpha$-$\gamma$ transformation.
% structural phase transition.
%, which is necessary for theoretical description of the transition.   

\end{abstract}

\maketitle

{\it Introduction}. Iron is one of the substances known from ancient times. 
%Presently, it is used in many technological applications, such as producing of various steels, memory devices, etc. 
Many technologically important applications of iron and its alloys, such as producing steels, are dealt with the structural transition between the \a phase with a body-centered cubic (bcc) lattice and the \g phase with a face-centered cubic (fcc) lattice. 
In pure iron this transition occurs in the paramagnetic region at 1185~K slightly above the Curie temperature of 1043~K. The theoretical description of this transition is important from both, fundamental and practical points of view.

%In last decades the theoretical studies of Fe were mainly performed using the density functional theory (DFT). 
%
The ground-state properties of $\alpha$ and $\gamma$ phases were extensively studied\cite{Fe_DFT} by the density functional theory (DFT) methods, in particular local density approximation (LDA) and generalized gradient approximation (GGA); the disordered local moment (DLM) approach~\cite{dlm_method} was applied to simulate the paramagnetic state by randomly distributed magnetic moments.
%
%\com{OTHER VERSION: }
%In the ground state, the energies of $\alpha$ and $\gamma$ phases were obtained within the density functional theory (DFT) methods, 
%in particular, local density approximation (LDA) or generalized gradient approximation (GGA) for the magnetically ordered state\cite{Fe_DFT} and disordered local moment (DLM) approach~\cite{dlm_method}, simulating the paramagnetic state by randomly distributed magnetic moments.
%
%
%
%, or special quasirandom structures method (SQS) \cite{SQS}, simulating average over some magnetic configuarations, in the non-magnetic state.
The energies of various phases were compared and the respective correct values of magnetic moments at zero temperature were obtained within these studies \cite{Fe_DFT,Okatov09,Zhang11}.
%,Mankovsky13}.
%-========
%The pure band structure methods allow to predict only the ground-state properties of iron. In particular, density functional theory (DFT) calculations of Fe within local density approximation (LDA) or generalized gradient approximations (GGA) resulted in a good description of magnetic moments and energies of magnetically ordered ground state~\cite{Fe_DFT}.
%, while the paramagnetic ones cannot be properly described by non-magnetic DFT calculations.
%
%For Fe this problem is often solved
%To describe energies of paramagnetic phases the DFT methods were combined with
%with Heisenberg model or 
%disordered local moment (DLM) approach~\cite{dlm_method} simulating the paramagnetic state by randomly distributed magnetic moments, or special quasirandom structures method (SQS) \cite{SQS}, simulating average over some magnetic configuarations, yielding correct description of the energy of paramagnetic phase \cite{Okatov09,Zhang11,Mankovsky13} in comparison with the ordered phases.
%
The combination of these methods with the Heisenberg model gave a possibility to treat magnetic correlations (also at finite temperature), and provided an accurate value for the Curie temperature of bcc Fe
%, but underestimated the effective local magnetic moment.
~\cite{DLM_Tc,Dusseldorf2}, its thermodynamic properties~\cite{Dusseldorf134,Gornostyrev}, and magnon-phonon coupling~\cite{Dusseldorf5}. 
%
%The latter approach 
%
%This combination was also used to study the structural transformations in iron as a function of temperature and carbon concentration~\cite{Gornostyrev}. 
%
This combination also resulted in an accurate value for the alpha-gamma transition temperature as a function of carbon concentration~\cite{Gornostyrev}.

%Another problem of DFT deals with considering Coulomb correlation effects, which are poorly treated in LDA and GGA. 
%
%This problem can be overcome by the so-called LDA+$U$ method~\cite{} employing static approximation for the on-site Coulomb interaction, which is, however, applicable only for systems with long-range magnetic order.
%
%Both the paramagnetic state and Coulomb correlations can be considered by a combination of DFT methods with dynamical mean-field theory~\cite{DMFT} (DMFT).
%
%This combination is usually called the LDA+DMFT method~\cite{LDA+DMFT} and at the moment is one of the most effective tools for strongly correlated transition metals and their compounds.
%
% Neglecting nonlocal correlation effects, the DMFT carefully takes into account the local spin dynamics. 

%At present 
Despite these successes, the described methods do not consider important local correlations in iron, and, therefore, do not provide a comprehensive view on the $\alpha$-$\gamma$ transition. To treat the effect of local correlations we apply in the present paper 
%A powerful tool for theoretical study the effect of electronic correlations in transition metals and their compounds is 
the  combination of dynamical mean-field theory (DMFT) \cite{DMFT} with density functional theory (DFT) methods, usually called LDA+DMFT \cite{LDA+DMFT}. 
%strongly correlated 
%
%However, the nonlocal correlation effects are neglected in conventional DMFT. 
%
%To include nonlocal correlations, different extensions, e.g., dynamic vertex
%approximation~\cite{DGA} and dual fermion approach~\cite{dual_fermions} have been developed.
%
%However, these appoaches are too computationally expensive to be applied to real multiorbital compounds at the moment. 
Previous studies by LDA+DMFT 
%were successful in describing spectral and magnetic  properties of iron ~\cite{Lichtenstein2001,Leonov_alpha-gamma,Leonov_phonons,OurAlpha0,OurGamma,Belozerov2013,Pourovskii,OurAlpha,Sangiovanni}. In particular, these studies 
allowed one to obtain the correct values of magnetic moments in $\alpha$ and $\gamma$ phases \cite{Lichtenstein2001,OurAlpha0,OurGamma,OurAlpha,Sangiovanni}, the linear behavior of the temperature dependence of the inverse local\cite{OurAlpha0,OurGamma,OurAlpha,Sangiovanni} ($\alpha$,$\gamma$ phases) and uniform magnetic~\cite{Lichtenstein2001,Belozerov2013,Pourovskii} ($\alpha$~phase) susceptibilities, and revealed the non-monotonic temperature dependence of inverse the uniform magnetic susceptibility in the $\gamma$ phase in a broad temperature range \cite{OurGamma}.
%, which agrees qualitatively with the experimental temperature dependence in the temperature range, where this phase exist in nature.
%
%One of the well-known problems, where nonlocal effects are expected to result in signicant corrections, is 
%the structural transition in iron between \a phase with body-centered cubic (bcc) lattice and \g phase with face-centered cubic (fcc) lattice.
%This transition occurs in the paramagnetic region at 1185~K slightly above the Curie temperature of 1043~K.
%
%

In most of these studies the Curie temperature of the $\alpha$ phase was found, however, to be substantially overestimated. 
As a result, the description of the magnetization~\cite{Lichtenstein2001}, the temperature of the \ag transition~\cite{Leonov_alpha-gamma}, and the phonon spectra~\cite{Leonov_phonons} was provided in units of the calculated Curie temperature. 
The overestimation of the Curie temperature mainly comes from the DMFT part and is due to using the approximate (density-density) form of the Coulomb interaction~\cite{Belozerov2013,Sangiovanni} and neglecting nonlocal correlations in DMFT.
%(should we really cite Lichtenstein here?)~\cite{Lichtenstein2001}.

To solve the former problem, we apply in the present study the spin-rotationally-invariant DMFT approach \cite{Belozerov2013}.
%============ Version 2 ============
%
% In particular, the effective local magnetic moments in \a and \g phases in the paramagnetic state~\cite{}
% were well captured.
%
%However, the Curie temperature of \a-Fe in most of these studies was found to be substantially overestimated. 
%
%As a result, the description of magnetization~\cite{Lichtenstein2001}, phonon spectra~\cite{Leonov_phonons} and \ag transition~\cite{Leonov_alpha-gamma} was provided in units of the calculated Curie temperature. 
%
%The overestimation of the Curie temperature mainly comes from the DMFT part and is due to neglecting nonlocal correlations~\cite{Lichtenstein2001} and using the approximate (density-density) form of Coulomb interaction~\cite{Belozerov2013}.
%
Although the nonlocal corrections to DMFT can be taken into account using, e.g., the dynamic vertex
approximation~\cite{DGA}, the dual fermion approach~\cite{dual_fermions}, or cluster methods~\cite{clusters},
%
%However, 
%
these approaches are too computationally expensive to be applied to real multiorbital compounds at the moment. 

For iron, the nonlocal degrees of freedom can be described within the effective Heisenberg model, which was combined previously with DFT approaches in Refs. \onlinecite{Dusseldorf2,Dusseldorf134,Dusseldorf5,Gornostyrev}. However, a derivation of this model from microscopic principles, and its combination with a treatment of local correlations within LDA+DMFT was not considered previously.
%The results previously obtained for energies of alpha and gamma phases in DFT (Alexander, please write), local moments in alpha and gamma phases from DMFT (I write), energies in DMFT (write together), frozen phonon calculations in DMFT (why do they reproduce correctly position of alpha-gamma transition, if nonlocal correlations are required)? 
In the present paper we address the microscopic derivation of an effective Heisenberg model in the presence of local moments and calculate the nonlocal correction to the energy of the $\alpha$ phase near the magnetic phase transition. We show that this correction is crucially important to compare the energies of $\alpha$ and $\gamma$ phases near the structural phase transition without adjustable parameters.  
%Our study has no adjustable parameters, the parameters of the Coulomb interaction ${U=4}$~eV and ${J=0.9}$~eV are taken from Ref.~\onlinecite{Belozerov_UJ} and were obtained by the constrained density functional theory (cDFT) calculations in the basis of $spd$ Wannier functions.
%
%\com{(In my opinion the last sentence is more suitable at the end of paragraph with Hamiltonian (1))}
%
%We revisit this problem by (i) accurate DFT calculations within the ELK code and appropriate construction of Wannier orbitals (ii) spin-rotation invariant DMFT (iii) inclusion of nonlocal correlations.
%Formulation of the model and summary of spin-invariant DMFT (Alexander, please write). 

Let us turn first to the {\it LDA+DMFT part}. We performed DFT calculations using the full-potential linearized augmented plane-wave method implemented in the ELK code supplemented by the Wannier function projection procedure (Exciting-plus code~\cite{elk}).
The Perdew-Burke-Ernzerhof form~\cite{pbe} of GGA was considered.
The calculations were carried out with the experimental lattice constant ${a=2.91}$~\AA\, for the $\alpha$ phase in the vicinity of the \ag transition~\cite{Basinski55}.
The lattice constant for the $\gamma$~phase was set to keep the experimental volume of the unit cell for the $\alpha$ phase.
The integration in the reciprocal space was performed using an 18$\times$18$\times$18 $\textbf{k}$-point mesh.
The convergence threshold for the total energy was set to $10^{-6}$~Ry.
From the converged DFT results we constructed effective Hamiltonians in the basis of Wannier functions, which were built as a projection of the original Kohn-Sham states to site-centered localized functions as described in Ref.~\onlinecite{Korotin08}, considering $spd$ states. This differentiates the present approach from that of Ref.~\onlinecite{Leonov_alpha-gamma}, where only $sd$ states were taken into account.
The difference of DFT total energies obtained in our non-magnetic calculations for \a and \g phases is 0.280~eV/at in agreement with previous DFT studies~\cite{Leonov_alpha-gamma,Friak01,Okatov11} resulting in values from 0.24 to 0.3~eV/at.
%
%Previous nonmagnetic DFT studies of \a and \g phases resulted in difference of ground state energies from 0.24 to 0.3~eV~\cite{Leonov_alpha-gamma,Friak01,Okatov11}. The obtained difference of DFT total energies is ${E^{\rm{DFT}}_\alpha-E^{\rm{DFT}}_\gamma=0.280}$~eV.  

The effect of local correlations is considered within the DMFT approach of Ref.~\onlinecite{Belozerov2013}, applied to the Hamiltonian
\begin{eqnarray}
\hat{H}_\textrm{DMFT}=\hat{H}_\textrm{DFT}^\textrm{WF}+\hat{H}_\textrm{Coul}-\hat{H}_\textrm{DC},
\end{eqnarray}
where $\hat{H}_\textrm{DFT}^\textrm{WF}$ is the effective Hamiltonian in the basis of Wannier functions constructed for states near the Fermi level,
$\hat{H}_\textrm{Coul}$ is the on-site Coulomb interaction Hamiltonian,
and $\hat{H}_\textrm{DC}$ is the double-counting correction.
This correction was considered in the fully localized limit and had the form ${\hat{H}_\textrm{DC}=\bar{U}(n_\textrm{DMFT}^\textrm{d}- 1/2)}$,
where $n_\textrm{DMFT}^\textrm{d}$ is the number of d electrons in DMFT, and $\bar{U}$ is the average Coulomb interaction in the $d$ shell. We choose the on-site Coulomb and Hund interaction parameters ${U\equiv F^0=4}$~eV and ${J_S\equiv (F^2+F^4)/14=0.9}$~eV, where $F^0$, $F^2$, and $F^4$ are the Slater integrals as obtained in Ref.~\onlinecite{Belozerov_UJ} by the constrained density functional theory (cDFT) in the basis of $spd$ Wannier functions.

\begin{figure}[t]
%\centering
% \vspace{-2.7cm} 
\includegraphics[clip=false,width=0.48\textwidth]{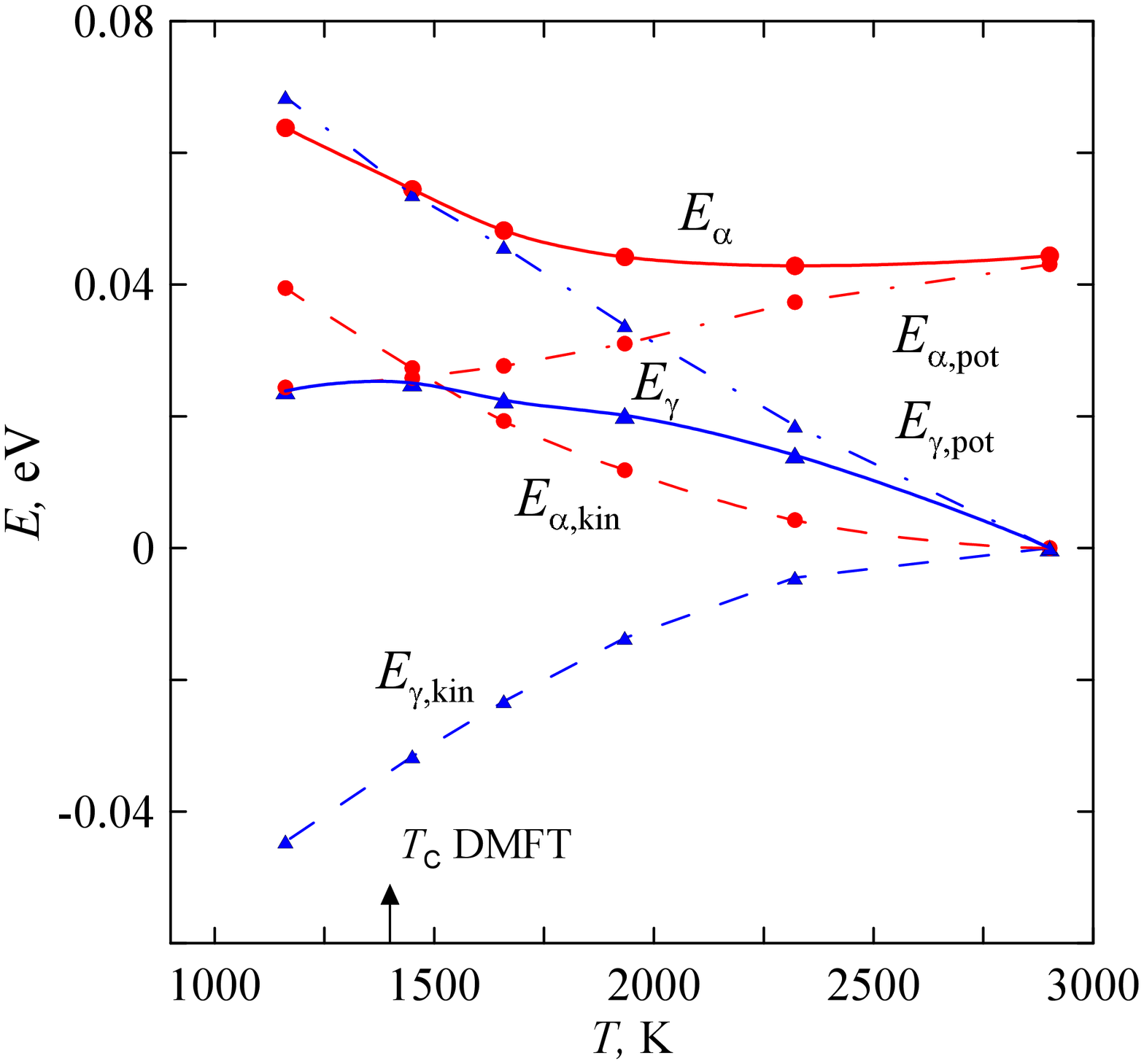}
\vspace{-0.45cm}
\caption{(Color online)
Temperature dependence of total (solid lines), kinetic (dashed lines), and Coulomb (dot-dashed lines) energies per atom obtained by LDA+DMFT for the $\alpha$ (red curves, circles), and the $\gamma$ (blue curves, triangles) phases of iron. The potential and kinetic energies of the $\alpha$($\gamma$) phase are shifted by $\pm E_{\alpha(\gamma),{\rm kin}}(T=2900{\rm K})$ for better view.  
\label{fig:model_susc}}
\end{figure}

%Let us consider DFT+DMFT results. 
%For the local moment and Curie temperature of the $\alpha$ phase we obtain $\mu^2 _{{\rm eff},\alpha}=2.7\mu_{\rm B}^2$, $T^{\alpha,{\rm DMFT}}_C=1400K$ in agreement with previous study\cite{Belozerov2013}.
From the uniform magnetic susceptibility of the $\alpha$ phase, we extracted the effective local  
moment ${\mu^2 _{{\rm eff},\alpha}=2.7\mu_{\rm B}^2}$ and the Curie temperature ${T^{\alpha,{\rm DMFT}}_C=1400}$~K in agreement with a previous study\cite{Belozerov2013}. As in this study, we expect that the Curie temperature is weakly dependent on Hubbard $U$ and is more sensitive to the Hund's coupling. 
The DMFT results for the energies are shown in Fig. 1. One can see that the energy of the $\alpha$ phase strongly increases with decreasing temperature. Looking at the partial contributions from kinetic and potential energies, one can see that the increase of the energy of the $\alpha$ phase with decreasing $T$ is due to the strong increase of the kinetic energy, while the potential energy expectedly decreases, reflecting the increase of instantaneous magnetic moment $\langle
\mathbf{S}_{i,\alpha}^{2}\rangle$ (Ref. \onlinecite{OurGamma}). 
%{\it Can one say that the increase of $E_{kin}$ was missed/underestimated in Lenonov's work due to incorrect/inaccurate Wannier projection etc?}. 
Although the energy of the $\gamma$ phase also increases with decreasing $T$, it saturates in the temperature range $1000-1500$~K. Moreover, inspection of kinetic and potential energies shows the opposite tendencies to those in the $\alpha$ phase: the mentioned increase of total energy upon cooling is provided by a strong increase of the {\it potential} energy, and a weaker decrease of the {\it kinetic} energy. The increase of potential energy reflects a decrease of instantaneous moment $\langle
\mathbf{S}_{i,\gamma}^{2}\rangle$ (Ref. \onlinecite{OurGamma}), and, compared to the opposite tendency of the $\alpha$ phase, provides a mechanism of stabilization of the $\alpha$ phase at low $T$. However, this mechanism is not the only contribution, and at the level of DMFT the $\gamma$ phase is "protected" by the respective decrease in kinetic energy in the $\gamma$ phase and its increase in the $\alpha$ phase. 
%(can we confirm this by direct calculation?)
%{\it Note: the increase of $E_{kin}$ in $\alpha$ phase is probably due to better coherence at low T, but why $E_{kin}$ decreases in $\gamma$ phase?}.   

%\section{Formulae for transition temperatures and correlation functions}

{\it Non-local corrections.} To calculate the nonlocal corrections to the Curie temperature and the
energy of the $\alpha$ phase, we treat the effect of local moments on the energy within the spin-fermion
model of Ref. \onlinecite{OurAlpha}, supplemented by a soft spin constraint, 
\begin{align}
\mathcal{S} &  =\sum_{ii',\nu_{n}\sigma ll^{\prime}}
c^\dagger_{il \sigma
}(\mathrm{i}\nu
_{n})\left[  \mathrm{i}\nu_{n}\delta_{ii^{\prime}}\delta_{ll^{\prime}}+H^{ii'}_{ll^{\prime}}-\Sigma_{ll^{\prime}}(\mathrm{i}\nu_{n})\right]%
\notag\\
&\times c_{i'l^{\prime}\sigma
}(\mathrm{i}\nu_{n})+\frac{1}{2}\sum_{i,\omega_{n}}\chi_{S}^{-1}(\mathrm{i}%
\omega_{n})\mathbf{S}_{i}(\mathrm{i}\omega_{n})\mathbf{S}_{i}(-\mathrm{i}%
\omega_{n})\nonumber \\
&+2J_K\sum_{i,\omega_{n}}\mathbf{S}_{i}(\mathrm{i}\omega_{n})\mathbf{s}%
_{i}(-\mathrm{i}\omega_{n})+\frac{1}{4}\sum_{i,\omega_n,\omega'_n}\lambda(\mathrm{i} \omega_n)\label{Seff}\\
&  \times \left\{  [\mathbf{S}_{i}(\mathrm{i}\omega_n)\mathbf{S}_{i}(-\mathrm{i}\omega_n+\mathrm{i}\omega'_n)]^2-\delta_{\omega'_n,0}\langle|\mathbf{S}_{i}(\mathrm{i}\omega_n)|^{2}\rangle
^{2}\right\},  \nonumber
\end{align}
where $i,l$ are site and orbital indices, $H=H^{\rm WF}_{\rm DFT}$, $\Sigma_{ll'}$ are the DMFT self-energies, $\mathbf{S}_i$ corresponds to the spin of the local-moment degrees of freedom,
$\mathbf{s}_i=\sum_{l\sigma \sigma'} c_{il\sigma}^{\dag} \mbox {\boldmath $\sigma $}_{\sigma \sigma'} c_{il\sigma'} $ to the spin of itinerant degrees of freedom, and $\mbox {\boldmath $\sigma $}$ are the Pauli matrices. The first and second lines in Eq. (\ref{Seff}) describe the propagation of itinerant electrons and the dynamics of the local moments, the third line corresponds to their interaction via Hund exchange $J_K\simeq (5/7)J_S$ in Kanamori parameterization, and the fourth line adds a spin constraint on the local moments, which restricts the size of the moment. This model can be considered as a simplified version of the multiorbital model, studied by LDA+DMFT, where the major effect of correlations -- formation of the local moments -- is incorporated in the local variables ${\bf S}$. The bare local moment
propagator $\chi_{S}^{-1}(\mathrm{i}\omega_{n})=4\mu_{B}^{2}\chi_{\text{loc}%
}^{-1}(i\omega_{n})+(J_K/\mu_{B})^{2}\overline{\chi}^{\mathrm{irr}}$ was
obtained in Ref. \onlinecite{OurAlpha}; $\chi_{\text{loc}}(i\omega_{n})$ is the
dynamic local susceptibility and  $\overline{\chi}^{\mathrm{irr}%
}\approx2\mu_{\rm B}^2/eV$ is the static local two-particle irreducible susceptibility in the $\alpha$ phase. 
%The reason for restricting static component is that it is was found as almost temperature independent in both, $\alpha$ and $\gamma$ phases. 
Decoupling the four-spin interaction in the soft constraint part in Eq. (\ref{Seff}) within bosonic mean-field theory (which implies neglecting critical fluctuations near $T_C^{\alpha}$), we obtain:
\begin{align}
\mathcal{S}&   =\frac{1}{2}\sum_{i,\omega_{n}}\left\{[\chi_{S}^{-1}(\mathrm{i}%
\omega_{n})+\Lambda(T,\mathrm{i} \omega_n )]\mathbf{S}_{i}(\mathrm{i}\omega_{n})\mathbf{S}%
_{i}(-\mathrm{i}\omega_{n})\right.\\
&  -
%\frac{1}{2}\sum_{i,\omega_n}
\left.\Lambda(T,\mathrm{i}\omega_n)\langle|\mathbf{S}_{i}(\mathrm{i} \omega_n)|%
^{2}\rangle\right\}+2J_K\sum_{i,\omega_{n}}\mathbf{S}_{i}(\mathrm{i}\omega_{n})\mathbf{s}%
_{i}(-\mathrm{i}\omega_{n})\nonumber\\
+&  \sum_{ii',\nu_{n}\sigma ll^{\prime}}
c^\dagger_{il \sigma
}(\mathrm{i}\nu
_{n})\left[  \mathrm{i}\nu_{n}\delta_{ii^{\prime}}\delta_{ll^{\prime}}+H^{ii'}_{ll^{\prime}}-\Sigma_{ll^{\prime}}(\mathrm{i}\nu_{n})\right]%
c_{i'l^{\prime}\sigma
}(\mathrm{i}\nu_{n}),\nonumber
\end{align}
where $\Lambda(T,\mathrm{i} \omega_n)=\lambda(\mathrm{i} \omega_n)\langle|\mathbf{S}_{i}(\mathrm{i} \omega_n)|^{2}\rangle.$ In the second
order in $J_K$ we find the corresponding effective model for spin degrees of freedom (cf. Ref. \onlinecite{OurAlpha})%
\begin{align}
\mathcal{S}_\textrm{eff} &  =\frac{1}{2}\sum_{\mathbf{q},\omega_{n}}\left[\chi^{-1}(\mathbf{q}%
,\mathrm{i}\omega_{n})\mathbf{S}_\mathbf{q}(\mathrm{i}\omega_{n}%
)\mathbf{S}_\mathbf{-q}(-\mathrm{i}\omega_{n})\right.\nonumber \\
&
\left.-
%\frac{1}{2}\sum_{q,\omega_{n}}
\Lambda(T, \mathrm{i} \omega_n)%
\langle|\mathbf{S}_\mathbf{q}(\mathrm{i}\omega_{n}%
)|^2\rangle\right],\label{S1}
\end{align}
where 
\begin{equation}
\chi^{-1}(\mathbf{q},\mathrm{i}\omega_{n}) =4\mu_{B}^{2}\chi_{\mathrm{loc}%
}^{-1}(\mathrm{i}\omega_{n})+\Lambda(T,\mathrm{i}\omega_n)-J_{\mathbf{q}},
\end{equation}
$J_{\mathbf{q}}=(J_K/\mu_{B})^{2}(\chi_{\mathbf{q}}^{\mathrm{irr}%
}-\overline{\chi}^{\mathrm{irr}})$ is the exchange interaction, $\chi_{\mathbf{q}}^{\mathrm{irr}%
}$ is the static two-particle irreducible susceptibility, which can be calculated as a bubble, constructed from itinerant Green functions\cite{OurAlpha}. The determination of the function $\Lambda(T,\mathrm{i} \omega_n)$ is a rather complicated problem, since it requires knowledge of the $\langle\mathbf{S}^2\rangle^2$ interaction potential in Eq. (\ref{Seff}). We fix its static component by the equality of the obtained static part of the on-site spin correlation function to that, obtained in DMFT $\mu^2_{\rm{eff},\alpha}=3T\chi_{\mathrm{loc}}(0)$; the latter is found to be almost temperature independent (contrary to the instantaneous moment $\langle \mathbf{S}_i ^2 \rangle$) in 
%both, $\alpha$ and $\gamma$ phases in 
a broad temperature range~\cite{OurGamma}. The corresponding condition reads
\begin{equation}
3T\sum_{\mathbf{q}}\frac{1}{\lambda_{0}-J_{\mathbf{q}}}=\frac{\mu
_{\mathrm{eff,}\alpha}^{2}}{4\mu_{\mathrm{B}}^{2}}, \label{Fix_mom}
\end{equation}
where $\lambda_{0}=4\mu_{B}^{2}%
\chi_{\mathrm{loc}}^{-1}(0)+\Lambda(T,0)$. The equation (\ref{Fix_mom}) is analogous to the one, obtained in the (static) spherical approximation to the classical Heisenberg model
\begin{equation}
H_{\rm Heis}=-\frac{1}{2}\sum_{ij}J_{ij}\mathbf{S}_{i}\mathbf{S}_{j}.\label{H}%
\end{equation}
Indeed, this model, treated in the spherical approximation, yields the action 
\begin{equation}
\mathcal{S}=\frac{1}{2}\sum\limits_{\mathbf{q}}(\lambda
_{0}-J_{\mathbf{q}})|\mathbf{S}_{\mathbf{q}}|^{2}-\frac{\lambda_{0}}{2}\langle
\mathbf{S}_{i}^{2}\rangle_{\rm Heis}\label{SSph1}%
\end{equation}
and the corresponding condition, Eq. (\ref{Fix_mom}), with ${\mu
_{\mathrm{eff,}\alpha}^{2}}={4\mu_{\mathrm{B}}^{2}}\langle
\mathbf{S}_{i}^{2}\rangle_{\rm Heis}$. Equation~(\ref{SSph1}) is also essentially equivalent to the static limit of Eq. (\ref{S1}) up to the local contribution, which does not depend on $\mathbf{S}_{\mathbf{q}}$.

\begin{figure}[b]
\centering
%\vspace{-2.8cm}
\includegraphics[width=8.5cm]{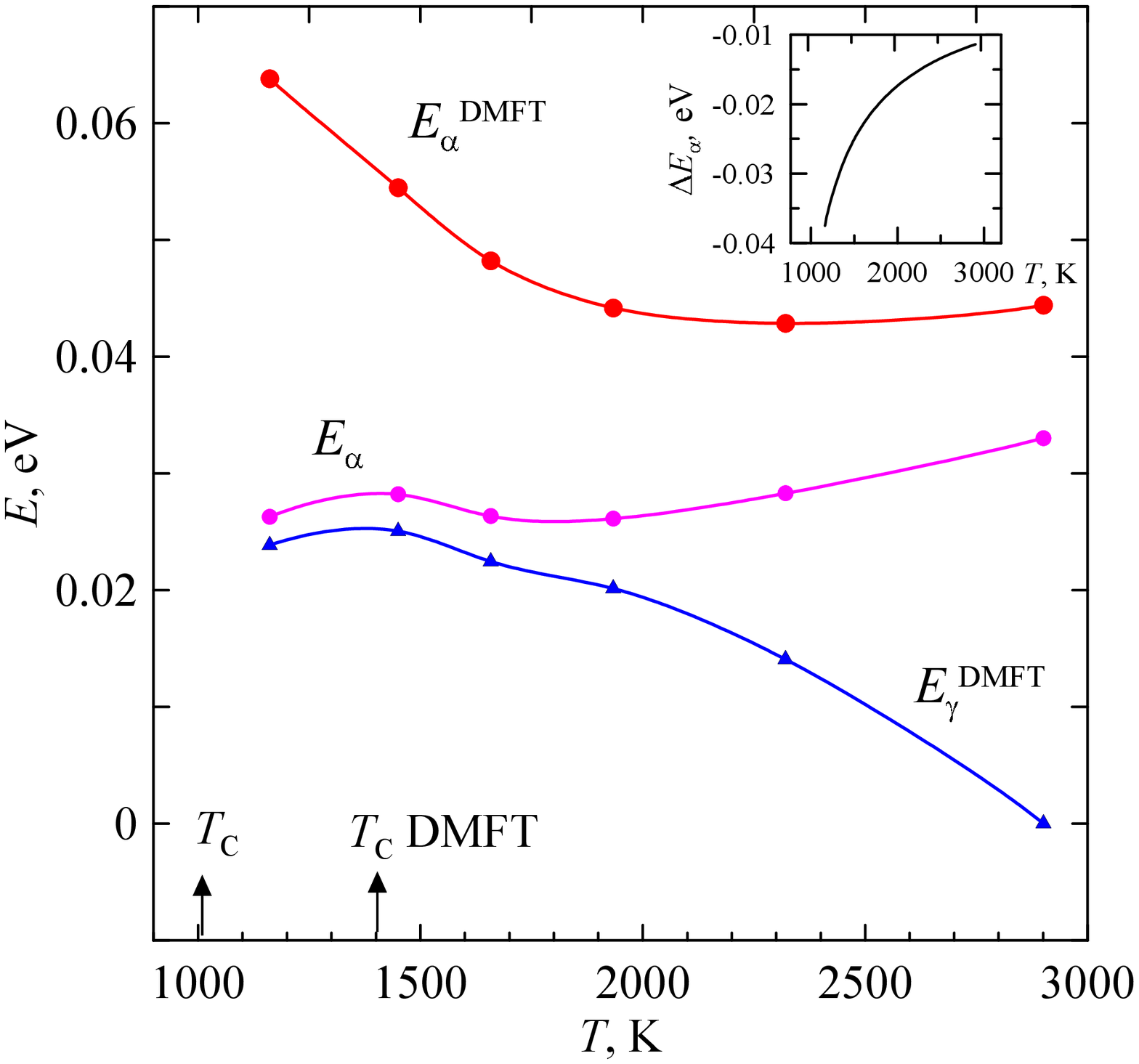}
\caption{(Color online)
Temperature dependence of total energies per atom of $\alpha$ (upper, red curve, circles) and $\gamma$ (blue curve, triangles) phases obtained by LDA+DMFT, and the energy of the $\alpha$ phase with the nonlocal correction (middle, magenta curve). Inset shows the nonlocal correction to the energy of the $\alpha$ phase, 
$\Delta E_\alpha=E_\alpha-E_\alpha^{\rm DMFT}$, calculated from Eq.~(\ref{eq:E_alpha}).
\label{fig:model_susc}}
\end{figure}

The Curie temperature is determined by vanishing of the gap of
the paramagnon spectrum, and in the same static approximation reads:%
\begin{equation}
3T^{\alpha}_{C}\sum_{\mathbf{q}}\frac{1}{J_{0}-J_{\mathbf{q}}}=\frac{\mu_{\mathrm{eff,}%
\alpha}^{2}}{4\mu_{\mathrm{B}}^{2}}.%
\end{equation}
%Therefore, we have $\Lambda(T_{C},0)=J_{0}-3(2\mu_{\mathrm{B}}/\mu_{\mathrm{eff,}\alpha})^{2}T_{C}.$ 
In the following we assume the nearest-neighbor approximation $J_{\bf q}=8J\cos(q_x/2)\cos(q_y/2)\cos(q_z/2)$, as justified in Refs. \onlinecite{Okatov11, OurAlpha}. %Although the exchange integral can be calculated from the DMFT results itself \cite{OurAlpha}, to avoid small inconsistencies, 
The exchange integral $J$ can be extracted from
the Curie temperature without nonlocal corrections (i.e. in DMFT, cf. Refs.
\onlinecite{OurAlpha,OurGamma}),
\begin{equation}
T_{\mathrm{C}}^{\alpha,{\rm DMFT}}=\frac{J_{0}}{3}\frac{\mu_{\mathrm{eff},\alpha}^{2}}%
{4\mu_{\mathrm{B}}^{2}}. \label{TcDMFT}
\end{equation}
%which we associate with the DMFT Curie temperature. 
Using this, we find
$T^{\alpha}_{\mathrm{C}}<T_{\mathrm{C}}^{\alpha,{\rm DMFT}}$ and $\Lambda(T^{\alpha}_{\mathrm{C}}%
,0)=3(2\mu_{\mathrm{B}}/\mu_{\mathrm{eff,}\alpha})^{2}(T_{\mathrm{C}}^{\alpha,{\rm DMFT}}-T^{\alpha}_{\mathrm{C}})>0.$
The corresponding nonlocal contribution to the energy of the $\alpha$ phase is obtained from the
Eq. (\ref{S1}) in the static approximation or from Eq. (\ref{SSph1}),
\begin{align}
E_{\alpha} &  =E^{\text{DMFT}}_{\alpha}-\frac{3T}{2}\sum_{\mathbf{q}}\frac{J_{\mathbf{q}}%
}{\lambda_{0}-J_{\mathbf{q}}}\nonumber\\
%&=E^{\text{DMFT}}_{\alpha}+\frac{3T}{2}-\frac{1}{2}\lambda_0 \frac{\mu_{\mathrm{eff},\alpha}^{2}}%
%{4\mu_{\mathrm{B}}^{2}}
%\nonumber
%\\
&  =E^{\text{DMFT}}_{\alpha}-\frac{\Lambda(T,0)}{2}\frac{\mu_{\mathrm{eff},\alpha}^{2}}%
{4\mu_{\mathrm{B}}^{2}}. \label{eq:E_alpha}
\end{align}
Since $0<\Lambda(T,0)<\Lambda(T_C^\alpha,0)$ at $T>T_C^\alpha$, the obtained correction is negative, decreasing the energy; the decrease is maximal at $T_C^\alpha$.  

In principle, the same calculation could be applied to obtain the nonlocal correction to the energy of the $\gamma$ phase. However, since the corresponding Neel temperature is much lower, than $T_C^{\alpha}$ (see, e.g., Ref. \onlinecite{OurGamma}), and pronounced corrections are obtained only in the vicinity of the magnetic transition temperature, we do not expect a substantial correction in that case.

Using Eq. (\ref{TcDMFT}) we find $J_0=0.20$~eV, which is close to the estimates of Refs.~\onlinecite{Okatov11} and \onlinecite{OurAlpha}. The corresponding Curie temperature with account of nonlocal correlations $T_C^{\alpha}=1005$~K is in a good agreement with the experimental data. 
The resulting temperature dependence of the energy of the $\alpha$ phase is shown in Fig. 2, together with the energies of $\alpha$ and $\gamma$ phases, obtained in DMFT. One can see, that
%introducing the nonlocal correction makes the energies of $\alpha$ and $\gamma$ phases very close in the vicinity of the $\alpha$-$\gamma$ transition.
the obtained nonlocal correction to the energy of the \a phase compensates the increase of its kinetic energy upon cooling and makes the energies of $\alpha$ and $\gamma$ phases very close in the vicinity of the $\alpha$-$\gamma$ transition. This demonstrates from one side, that the non--local corrections are crucially important for the description of this transition, and from the other side, the proposed methods are capable of describing adequately the effect of nonlocal correlations. 
%One can expect, that treating free energies, which is more numerically expensive and beyond the scope of present paper, will allow one to describe the $\alpha$-$\gamma$ transition.
The description of the alpha-gamma transition can be further improved by, e.g., using a more advanced rotationally-invariant quantum impurity solver than in our study (see, e.g., Ref. \onlinecite{Sangiovanni}). Another improvement can be made by considering free energies. However, at the moment such calculations are too computationally expensive and beyond the scope of the present paper.

We note that our results considerably differ from those of Leonov \textit{et al.}~\cite{Leonov_alpha-gamma}, where the $\alpha$-$\gamma$ transition was captured by LDA+DMFT with density-density interaction in units of the overestimated Curie temperature (1600~K). Aside from using the rotationally-invariant interaction and considering the absolute temperature dependencies, there are some computational details that differ in our study. In particular, (i) we use the all-electron full-potential LAPW method implemented in the ELK code resulting in a difference of DFT total energies of 0.280~eV, while Leonov \textit{et al.} used the pseudopotential Quantum ESPRESSO package leading to 0.244~eV. (ii) We use the $spd$ Wannier function basis, while only $sd$ states were included by Leonov \textit{et al.} (iii) We use Hubbard ${U=4}$~eV, while a much smaller value ${U=1.8}$~eV was employed by Leonov \textit{et al.}

%Although it is difficult to directly compare our results with those of Leonov \textit{et al.}, there are some discrepancies which become clear when considering the density-density interaction. In this case 
Within the above-mentioned methods we found the energy of the $\gamma$ phase to be at least 0.044~eV below the $\alpha$ phase in the paramagnetic region for the density-density interaction, yielding a Curie temperature $\sim$2150~K, which is larger than the 1600~K value obtained by Leonov \textit{et al.} and is close to 1900~K obtained by Lichtenstein \textit{et al.}~\cite{Lichtenstein2001} Previous LDA+DMFT studies of iron indicated that the Curie temperature is weakly dependent on Hubbard~$U$ (Ref.~\onlinecite{Belozerov_UJ}). Therefore we expect that this discrepancy is mainly due to different Wanner function basis. This is supported by the fact that the DMFT calculations by Lichtenstein \textit{et al.} were performed with $3d$, $4s$, and $4p$ states included in the basis set (not Wannier functions). Since the total energy is a subtle quantity, we suppose that the discrepancy between our and Leonov \textit{et al.} results can be further influenced by the above mentioned computational differences (i)-(iii), but consider our calculation to be more accurate in these respects. To shed light on this point, further studies are required.

{\it In conclusion}, we have presented a method to evaluate the nonlocal correction to the Curie temperature and energy, obtained in DMFT in the presence of local moments by deriving the spherical approximation results for the effective Heisenberg model from the spin-fermion model. We have shown that the obtained results yield the energies of $\alpha$ and $\gamma$ phases, which are very close in the vicinity of the $\alpha$-$\gamma$ transition, which is necessary to describe the structural phase transition in iron.

The work was supported by the grant of the Russian Science Foundation (project no. 14-22-00004).


\begin{thebibliography}{9}                                                                                       
\bibitem{Fe_DFT} 
  D. J. Singh, W. E. Pickett, and H. Krakauer, Phys. Rev. B \textbf{43}, 11628 (1991); 
  L. Stixrude, R. E. Cohen, and D. J. Singh, \textit{ibid.} \textbf{50}, 6442  (1994);
  E. G. Moroni, G. Kresse, J. Hafner, and J. Furthmuller, \textit{ibid.} \textbf{56}, 15629 (1997). 

\bibitem{dlm_method} B. L. Gyorffy, A. J. Pindor, J. Staunton, G. M. Stocks, and H. Winter, J. Phys. F: Met. Phys. \textbf{15}, 1337 (1985). 

% \bibitem{SQS} A. Zunger, S.-H. Wei, L. G. Ferreira, and J. E. Bernard, Phys. Rev. Lett. {\bf 65}, 353 (1990).

\bibitem{Okatov09} S. V. Okatov, A. R. Kuznetsov, Yu. N. Gornostyrev, V. N. Urtsev, and M. I. Katsnelson, Phys. Rev. B \textbf{79}, 094111 (2009).

\bibitem{Zhang11} H. Zhang, B. Johansson, and L. Vitos, Phys. Rev. B \textbf{84}, 140411(R) (2011).

%\bibitem{Mankovsky13} S. Mankovsky, S. Polesya, H. Ebert, W. Bensch, O. Mathon, S. Pascarelli, and J. Minar, Phys. Rev. B \textbf{88}, 184108 (2013).

\bibitem{DLM_Tc} J. B. Staunton and B. L. Gyorffy, Phys. Rev. Lett. \textbf{69}, 371 (1992).
  
  % DFT for FM bcc Fe + Heisenberg model in MF and RPA  
\bibitem{Dusseldorf2} F. K\"ormann, A. Dick, T. Hickel, and J. Neugebauer, Phys. Rev. B {\bf 79}, 184406 (2009).
  
% DFT for FM bcc Fe + Heisenberg model in MF and RPA  
\bibitem{Dusseldorf134}
  F. K\"ormann, A. Dick, B. Grabowski, B. Hallstedt, T. Hickel, and J. Neugebauer, Phys. Rev. B {\bf 78}, 033102 (2008);
  F. K\"ormann, A. Dick, T. Hickel, and J. Neugebauer, Phys. Rev. B {\bf 81}, 134425 (2010);
  T Hickel, B Grabowski1, F Körmann and J Neugebauer,  J. Phys.: Condens. Matter {\bf 24}, 053202 (2012). 

\bibitem{Gornostyrev} I. K. Razumov, D. V. Boukhvalov, M. V. Petrik, V. N. Urtsev, A. V. Shmakov, M. I. Katsnelson, Yu. N. Gornostyrev, Phys. Rev. B \textbf{90}, 094101 (2014).

% DFT + Heisenberg
\bibitem{Dusseldorf5} F. K\"ormann, B. Grabowski, B. Dutta, T. Hickel, L. Mauger, B. Fultz, and J. Neugebauer, Phys. Rev. Lett. {\bf 113}, 165503 (2014).

\bibitem{DMFT}
  W. Metzner and D. Vollhardt, Phys. Rev. Lett. \textbf{62}, 324 (1989);
  A. Georges, G. Kotliar, W. Krauth and M. J. Rozenberg, Rev. Mod. Phys. \textbf{68}, 13 (1996).

\bibitem{LDA+DMFT} 
  V. I. Anisimov, A. I. Poteryaev, M. A. Korotin, A. O. Anokhin, and G. Kotliar, J. Phys.: Condens. Matter \textbf{9}, 7359 (1997); 
  A. I. Lichtenstein and M. I. Katsnelson, Phys. Rev. B \textbf{57}, 6884 (1998).

\bibitem{Lichtenstein2001}
  A. I. Lichtenstein, M. I. Katsnelson, and G. Kotliar, Phys. Rev. Lett. \textbf{87}, 067205 (2001).


\bibitem{OurAlpha0} A. A. Katanin, A. I. Poteryaev, A. V. Efremov, A. O. Shorikov, S. L. Skornyakov, M. A. Korotin, V. I. Anisimov, Phys. Rev. B 81, 045117 (2010).

\bibitem{OurGamma}P. A. Igoshev, A. V. Efremov, A. I. Poteryaev, A. A. Katanin, V. I. Anisimov, Phys. Rev. B \textbf{88}, 155120 (2013).

\bibitem{OurAlpha}P. A. Igoshev, A. V. Efremov, A. A. Katanin, Phys. Rev. B \textbf{91}, 195123 (2015).

\bibitem{Sangiovanni} A. Hausoel, M. Karolak, E. Sasioglu, A. Lichtenstein, K. Held, A. Katanin, A. Toschi,
and G. Sangiovanni (in preparation). 

\bibitem{Belozerov2013}
  A. S. Belozerov, I. Leonov and V. I. Anisimov, Phys. Rev. B \textbf{87}, 125138 (2013).

\bibitem{Pourovskii} L. V. Pourovskii, T. Miyake, S. I. Simak, A. V. Ruban, L. Dubrovinsky, I. A. Abrikosov, Phys. Rev. B 87, 115130 (2013).

\bibitem{Leonov_alpha-gamma}
  I. Leonov, A. I. Poteryaev, V. I. Anisimov, and D. Vollhardt, Phys. Rev. Lett. \textbf{106}, 106405 (2011).

\bibitem{Leonov_phonons}
  I. Leonov, A. I. Poteryaev, V. I. Anisimov, and D. Vollhardt, Phys. Rev. B \textbf{85}, 020401(R) (2012);
  I. Leonov, A. I. Poteryaev, Yu. N. Gornostyrev, M. I. Katsnelson, V. I. Anisimov and D. Vollhardt, Scientific Rep. \textbf{4}, 5585 (2014).

\bibitem{DGA}
  A. Toschi, A. A. Katanin, and K. Held, Phys. Rev. B \textbf{75}, 045118 (2007);
  A. Toschi, G. Rohringer, A. A. Katanin, K. Held, Ann. der Phys. \textbf{523}, 698 (2011).

\bibitem{dual_fermions} 
  A. N. Rubtsov, M. I. Katsnelson, A. I. Lichtenstein, A. Georges, Phys.Rev. B \textbf{79}, 045133 (2009).

\bibitem{clusters}
  M. H. Hettler, A. N. Tahvildar-Zadeh, M. Jarrell, T. Pruschke, and H. R. Krishnamurthy, Phys. Rev. B \textbf{58}, R7475(R) (1998);
  A. I. Lichtenstein and M. I. Katsnelson, Phys. Rev. B \textbf{62}, R9283 (2000); 
  M. H. Hettler, M. Mukherjee, M. Jarrell, and H. R. Krishnamurthy, Phys. Rev. B \textbf{61}, 12739 (2000);
  G. Kotliar, S. Y. Savrasov, G. Palsson, and G. Biroli, Phys. Rev. Lett. \textbf{87}, 186401 (2001).

\bibitem{elk}  http://elk.sourceforge.net/

\bibitem{pbe} J. P. Perdew, K. Burke, and M. Ernzerhof,  Phys. Rev. Lett. \textbf{77}, 3865 (1996).

\bibitem{Basinski55} Z. S. Basinski, W. Hume-Rothery, and A. L. Sutton, Proc. R. Soc. London, Ser. A \textbf{229}, 459 (1955).

\bibitem{Korotin08} Dm. Korotin, A. V. Kozhevnikov, S. L. Skornyakov, I. Leonov, N. Binggeli, V. I. Anisimov, and G. Trimarchi,  Eur. Phys. J. B \textbf{65}, 91  (2008).

\bibitem{Friak01}
  M. Friak, M. Sob, and V. Vitek, Phys.Rev. B \textbf{63}, 052405 (2001).

\bibitem{Okatov11} S. V. Okatov, Yu. N. Gornostyrev, A. I. Lichtenstein, and M. I. Katsnelson, Phys. Rev. B {\bf 84}, 214422 (2011).

\bibitem{Belozerov_UJ} A. S. Belozerov and V. I. Anisimov,  J. Phys.: Condens. Matter \textbf{26}, 375601 (2014).

\end{thebibliography}
\end{document}